\documentclass[11pt,a4paper]{article}
\pdfoutput=1
\usepackage{jcappub}
\graphicspath{{images/}}
\newcommand{\myat}{{\fontfamily{pag}\selectfont @}}
\usepackage{comment}
\usepackage{amsmath,bm,accents}
\usepackage{bm,accents}
\usepackage{booktabs}
\usepackage{multirow}
\makeatletter
\newcommand\accbm[1]{\use@mathgroup{\M@OMS}{7}{\bm{#1}}}
\makeatother
\title{\boldmath Axion-like particle searches with MeerKAT and SKA}

\author[a,1]{Ahmed Ayad\note{Corresponding author.}}
\author[a]{and Geoff Beck}
\affiliation[a]{School of Physics, University of the Witwatersrand,\\Private Bag 3, WITS-2050, Johannesburg, South Africa}

\emailAdd{ahmed\myat aims.edu.gh}
\emailAdd{Geoffrey.Beck\myat wits.ac.za }

\abstract{In the past few years, the search for axion-like particles (ALPs) has grown significantly due to their potential to account for the total abundance of the cold dark matter (CDM) in the universe. The coupling between ALPs and photons allows the spontaneous decay of ALPs into pairs of photons. For ALPs condensed in CDM halos around galaxies, the stimulated decay of ALPs is also possible. In this work, we examine the detectability of the radio emissions produced from this process with forthcoming radio telescopes such as the Square Kilometer Array (SKA) and MeerKAT. Our results, using recent more realistic sensitivity estimates, show that previous non-observation upper-limits projected for the SKA were highly optimistic, with the limits from dwarf galaxy observations being weakened by an order of magnitude at least. Notably, our results also depend far more strongly on ALP mass than previously, due to the inclusion of frequency dependent degradation effects. We show that the strongest potential environment to probe ALPs is nearby radio galaxies (due to the strong photon enhancement factor). In addition, with the use of a visibility taper, ALPs in the mass range of $4.96 \times 10^{-7} \ \text{--} \ 1.04 \times 10^{-4} \ {\rm eV}$ would have non-observation upper limits on the ALP-photon $g_{a\gamma}$ in the range of $1.83 \times 10^{-12} \ \text{--} \ 7.69 \times 10^{-10} \ {\rm GeV}^{-1}$ with SKA. MeerKAT can only produce limits similar to the CAST experiment within 50 hours of observation. Finally, we demonstrate that magnetic conversion of CDM ALPs to photons, in galactic magnetic fields, is highly sub-dominant, even to spontaneous decay.}

\arxivnumber{2010.05773}
\begin{document}
\maketitle
\flushbottom

\section{Introduction} \label{sec.1}

The existence of stable cold dark matter (CDM) is established based on many astrophysical and cosmological observations such as the cosmic microwave background (CMB), large scale structure, galactic rotation curves, and beyond~\cite{komatsu2011seven, peebles2015dark}. In particular, the Planck 2018 data predicts that the visible universe contains non-baryonic dark matter (DM), the abundance of which is estimated to be around five times greater than that of ordinary baryonic matter~\cite{aghanim2018planck}. However, the nature and properties of the DM remain one of the most pressing challenges in cosmology and particle physics. Although there are many suggested candidates for the DM content, there is currently no evidence strongly favoring any of them. The search for a broad class of light, pseudo-scalar candidates such as axions or, more generally, axion-like particles (ALPs) is one of the most promising approaches to understand the nature of the DM content~\cite{abbott1983cosmological, preskill1983cosmology, dine1983not, arias2012wispy}. The typical axions~\cite{peccei1977cp, weinberg1978new} are identified in the Peccei–Quinn (PQ) mechanism as pseudo-Nambu-Goldstone bosons that appear after the spontaneous breaking of the PQ symmetry. This being introduced to explain the absence of CP-violation for the strong interaction, which is theoretically obvious from the Lagrangian of quantum chromodynamics (QCD), for a review see reference~\cite{peccei2008strong}. Furthermore, many extensions of the standard model of particle physics (SM), including string theory models, generalized this concept to predict ALPs, which may arise as pseudo-scalar particles from the breaking of various global symmetries~\cite{arvanitaki2010string, cicoli2012type, anselm1982second}. In particular, ALPs naturally arise from string theory due to the compactification of extra dimensions~\cite{svrcek2006axions, witten1984some}. The phenomenology of ALPs is similar to that of QCD axions, characterized by their coupling with two photons. The main difference between them is that the QCD axion coupling parameter is determined by the axion mass; however, this is not necessarily the case for ALPs. Although the masses of the generic ALPs are expected theoretically to be very tiny, nowadays they are accounted among the leading candidates to compose a major part, if not all, of the DM content of the universe~\cite{abbott1983cosmological, ringwald2012exploring, marsh2016axion}. This is motivated by the theoretical predictions for their properties that are determined by their low mass and very weak interactions with the SM particles, see reference~\cite{asztalos2006searches}. It is worth noting here that it is believed that the main mechanisms to generate the ALP DM in the early universe are the misalignment mechanism and the decay of strings and domain walls~\cite{chang1998studies}. The abundance of the produced ALPs by these mechanisms is suitable to account for all the DM content~\cite{marsh2016axion}.

The coupling between ALPs and photons is possible due to the ALP-two-photon interaction vertex~\cite{sikivie1983experimental}. This ALP-photon coupling allows for the Primakoff conversion between ALPs and photons in the presence of an external magnetic field (see e.g. ~\cite{sigl2017astrophysical, huang2018radio, hook2018radio, ayad2020probing,ayad2019phenomenology}), as well as for the radiative decay of the ALPs into pairs of photons (see e.g. ~\cite{caputo2019detecting, caputo2018looking, ayad2020potential}). These two processes provide the theoretical basis for the majority of the recent mechanisms to search for ALPs in both laboratory experiments and astrophysical environments. Interactions with electrons and hadrons are also possible in the literature for the QCD axion. However, the main focus lies on the effect of ALP-photon coupling because of two reasons. The first is that the coupling with photons is a characteristic linking the QCD axion and generic ALPs. The second is due to the fact that the photon is the only SM particle that we are precisely confident is lighter than the generic DM ALPs~\cite{alonso2020wondrous}. However, increasing the effects of other couplings such as the self-couplings is doable through modifications to the standard ALPs~\cite{alonso2020wondrous}. Although the effects of the ALPs self-interactions remain very weak, they can become important in the thermal equilibrium of ALP condensates.

An essential feature in the context of ALP decay is the fact that ALPs are identical bosons; their very low mass indicates that their density and occupation numbers can be very high. Therefore, it is suggested that ALPs may form a Bose-Einstein condensate (BEC) with only short-range order~\cite{sikivie2009bose, davidson2013bose, davidson2015axions}. This ALP BEC can then thermalize through gravitational attraction and self-interactions to spatially localized clumps~\cite{chang1998studies, sikivie2009bose, erken2012cosmic}. By clump, we mean gravitationally bound structures composed of axions or ALPs, which may present in the dark matter halos around many galaxies; for more detail about the properties of these BEC clumps, see references~\cite{schiappacasse2018analysis, hertzberg2018scalar}. Of particular importance here is that the system in such high occupancy cases is well described by a classical field~\cite{guth2015dark}. Indeed, the spontaneous decay rate of ALPs is very small as a result of both their very low mass and their very weak coupling. However, the rapid expansion of the early universe would lead to an extremely homogeneous and coherent initial state of the ALP field. Hence, all ALPs are in the same state with very high occupation number. Therefore, the stimulated decay of ALPs into pairs of photons with a very high rate is very likely in the presence of an electromagnetic background with a certain frequency. During this process, the electromagnetic wave will be greatly enhanced and Bose enhancement effects seem plausible. In contrast, in a scenario of an empty and non-expanding universe, the resulting stimulated emission would induce extremely rapid decay of ALPs into photons, invalidating most of the interesting parameter space~\cite{alonso2020wondrous}. Nevertheless, the rapid expansion of the early universe and the plasma effects can disrupt this extremely fast process. These effects are crucial, as they allow ALPs to account for all the cosmological DM in the present epoch, while still permitting significant stimulated decay.

In the past years, there have been many proposals to detect signals from ALPs DM using radio telescopes, most of which were based on the Primakoff conversion of ALPs into photons~\cite{caputo2018looking, caputo2019detecting, Berezhiani:1989fu}. Recently it was shown that the stimulated decay of ALPs in the astrophysical environments may also generate radio signals comparable with that of the Primakoff conversion~\cite{caputo2018looking, sigl2017astrophysical}. In this work, we estimate the radio emissions that can be produced from the stimulated decays of ALPs into photons in particular astrophysical fields. Then, we compare these fluxes with the sensitivities of the next-generation radio telescopes such as the Square Kilometer Array (SKA)~\cite{dewdney2009square} and the MeerKAT~\cite{jonas2009meerkat} telescopes. Further, we explore the potential of these radio telescopes to probe new regions of the mass-coupling parameter space of ALPs. To this purpose, we extend previous works~\cite{caputo2018looking, caputo2019detecting, hertzberg2020merger} by considering the MeerKAT telescope and not just the SKA telescope. Additionally, we make use of realistic sensitivities for both instruments studied to derive more accurate potential constraints on the ALP-photon coupling. The results presented in this work argue that the sensitivity of radio telescope experiments to ALP stimulated decay signals has been quite strongly over-estimated. In particular, we show that SKA non-observation limits from dwarf galaxies are at least an order of magnitude weaker than those from~\cite{caputo2019detecting}, in addition to having much stronger ALP mass dependence than previously reported~\cite{caputo2018looking,caputo2019detecting}. This is due to the fact that we use more realistic SKA sensitivities, drawn from~\cite{braun2019anticipated}, that include frequency dependent degradation effects as well as specifically simulating line emission sensitivities. Our results indicate that the most promising environments are nearby radio galaxies, where the strong Bose enhancement outweighs the usual advantages of nearness and density enjoyed by dwarf galaxies. The resulting radio galaxy limits are similar to~\cite{caputo2019detecting} at low ALP masses, however, they weaken greatly with increasing mass. In order to allow radio telescopes to provide complementary probes to experiments like ALPS-II~\cite{bahre2013any} and IAXO~\cite{armengaud2019physics}, we explore the potential of a visibility taper. This tapering approach has been used for indirect radio detection of dark matter in previous work~\cite{regis2017dark}, and it is able to improve our limits on the ALP-photon coupling by about an order of magnitude. Finally, we determine that MeerKAT can only feasibly probe a region of parameter space similar to CAST~\cite{anastassopoulos2017new}.

The outline of this paper is as follows. The discussion in the previous paragraphs sets the plan for the rest of the manuscript. In section~\ref{sec.2}, we review the theoretical formalism for the spontaneous and stimulated decay of ALPs, as well as magnetic conversion processes. In section~\ref{sec.4}, we present our means to determine the surface brightness that can result from the stimulated decay of ALPs in DM halos. Next in section~\ref{sec.5}, we discuss the sensitivity of the near-future SKA and MeerKAT radio telescopes. In section~\ref{sec.6}, we present our results for the potential of the near-future SKA and MeerKAT radio telescopes to detect a radio signature produced from this process. Finally, our conclusion is provided in section~\ref{sec.7}.

\section{Interactions between ALPs and photons} \label{sec.2}

Most of the phenomenological implications of ALPs are due to their feeble interactions with the SM particles that take place through the Lagrangian~\cite{sikivie1983experimental, raffelt1988mixing, anselm1988experimental}
\begin{equation} \label{eq.2.1}
{\rm \ell}_{\rm a} = \frac{1}{2} \partial_{\mu} a \partial^{\mu} a - V(a) - \frac{1}{4} {\rm F}_{\mu \nu} \tilde{{\rm F}}^{\mu \nu} - \frac{1}{4} g_{a\gamma} a {\rm F}_{\mu \nu} \tilde{{\rm F}}^{\mu \nu} \:,
\end{equation}
where $g_{a\gamma}$ is the ALP-photon coupling parameter with dimension of inverse energy, ${\rm F}_{\mu \nu}$ and $\tilde{{\rm F}}^{\mu \nu}$ represent the electromagnetic field tensor and its dual respectively, and $a$ denotes the ALP field. The specific form of the potential $V(a)$ is model dependent. For the QCD axion, it results from non-perturbative QCD effects associated with instantons, which are non-trivial to compute with accuracy. For a more simple and general case, the ALPs potential can be written in terms of the ALP mass $m_{\rm a}$ and the energy scale of PQ symmetry breaking $f_{\rm a}$ in the simple form
\begin{equation} \label{eq.2.2}
V(a) = m_{\rm a}^2 F_{\rm a}^2 \left[ 1- \cos \left( \frac{a}{F_{\rm a}} \right) \right] \:.
\end{equation}
Since we shall only be interested in the non-relativistic regime for ALPs, we focus on very small field configurations $a \ll F_{\rm a}$. However, when ALP densities are high, we cannot neglect their self-interactions. In this case, the potential can be expanded as a Taylor series with the dominant terms
\begin{equation} \label{eq.2.3}
V(a) \approx \frac{1}{2} m_{\rm a}^2 a^2 - \frac{1}{4 !} \frac{m_{\rm a}^2}{F_{\rm a}^4} a^4+ \dots \:.
\end{equation}
Note that the specific values of the ALPs mass $m_{\rm a}$ and the energy scale $F_{\rm a}$ are model dependent. The coupling strength of ALP to two-photons can be given by the relation~\cite{marsh2016axion}
\begin{equation} \label{eq.2.4}
g_{a\gamma} = \frac{\alpha_{\rm em}}{2 \pi F_{\rm a}} \ C_{\rm a\gamma} \:,
\end{equation}
where $\alpha_{\rm em}$ is the electromagnetic fine-structure constant, and $C_{\rm a\gamma}$ is a dimensionless model-dependent coupling parameter, usually thought to be of order unity.

\subsection{Spontaneous decay rate of ALPs} \label{sec.2.1}

The ALP-two-photon interaction vertex allows for the radiative decay of the ALPs into pairs of photons. The spontaneous decay rate of an ALP with mass $m_{\rm a}$ in vacuum into pair of photons, each with a frequency $\nu=m_{\rm a} / 4\pi$, can be expressed from the usual perturbation theory calculations in term of its mass and ALP-photon coupling $g_{a\gamma}$~\cite{kelley2017radio}. The lifetime of ALPs can be given by the inverse of their decay rate as 
\begin{equation} \label{eq.2.5}
\tau_{\rm a} \equiv {\rm \Gamma}_{\rm pert}^{-1} = \frac{64 \pi}{m_{\rm a}^3 g_{a\gamma}^2} \:.
\end{equation}
This form is specific to QCD axions, and not for general ALPs. However, we can still use it as a reasonable estimation. For the typical QCD axions with mass $m_{\rm a} \sim 10^{-6} \ \rm{eV}$ and coupling with photons $g_{a\gamma} \sim 10^{-12} \ {\rm GeV}^{-1}$, we can insert these values into equation~\eqref{eq.2.4} and evaluate the perturbative decay time for general ALPs. This is found to be about $\sim 1.32 \times 10^{47} \ {\rm s}$. This lifetime for ALPs is much larger than the present age of the universe $\sim 4.3 \times 10^{17} \ {\rm s}$. Therefore ALPs seem to be super stable on the cosmic scale, and perhaps this is the main reason for neglecting the ALP decay in the literature. According to this scenario, the spontaneous decay of ALPs cannot be responsible for producing any observable signal that can be detectable by the current or the near-future radio telescopes. 

\subsection{Stimulated decay rate of ALPs} \label{sec.2.2}

Being identical bosons, ALPs are capable of forming a Bose-Einstein Condensate (BEC) with short-range order~\cite{sikivie2009bose, davidson2013bose, davidson2015axions,chang1998studies,schiappacasse2018analysis,hertzberg2018scalar,guth2015dark}. This ALP BEC can then form spatially localized clumps through gravitational attraction and self-interactions. These clumps are expected to have sufficiently large occupation numbers that the ALPs can be treated as a classical field~\cite{hertzberg2018dark}. In this circumstance, the decay of ALPs into two photons can be powerfully enhanced if the final state also has a large occupation number. For CDM-type ALPs this can be provided by various radio backgrounds, such as the CMB. Thus, given the very low mass of the ALPs, the stimulated emission in an empty and non-expanding universe would induce ALPs to decay very rapidly into photons, nullifying most of the interesting parameter space. The decay time obtained in~\cite{alonso2020wondrous} from the equation of motion for ALP condensates is $\sim 10^{-7} \ {\rm s}$, which is dramatically small compared to the perturbative decay time $\sim 10^{47} \ {\rm s}$. Indeed this scenario is problematic for ALP CDM scenarios, as it cannot lead to ALPs stable enough to account for the dark matter content of the universe. However, it is argued in~\cite{abbott1983cosmological, preskill1983cosmology, dine1983not,alonso2020wondrous} that there are two effects that can reduce the rate of the stimulated decay of ALPs into photons. The rapid expansion of the early universe redshifts the target photon population and, therefore, can discourage enhanced decay rates for ALPs. In addition to the expansion of the universe, plasma effects are crucial, as they modify the photon's propagation and prevent the early decay of ALPs. Inside the plasma, photons have an effective mass that kinematically forbids the decay of lighter particles including ALPs. Consequently, the redshifting of the decay products due to the expansion of the universe as well as the effective plasma mass of the photon can prevent the extremely fast decay rate of ALPs into photons. 

In this work we will treat ALPs as having an effective decay rate that can be expressed as~\cite{caputo2019detecting}
\begin{equation} \label{eq.2.8}
{\rm \Gamma}_{\rm eff} = {\rm \Gamma}_{\rm pert} (1+2 f_{\gamma}) \:,
\end{equation}
where $f_{\gamma}$ indicates the photon occupation number and the term between the brackets represents the stimulation factor due to the background photon radiation. It seems to be clear that only in the limits of very low photon occupation, $f_{\gamma} \ll 1$, will the effective decay rate be identical to the spontaneous decay rate ${\rm \Gamma}_{\rm pert}$ given by equation~\eqref{eq.2.5}. In addition, we will note that ALP decay will only be possible for masses such that produced photons are above the plasma frequency of their environment. This is given by
\begin{equation} \label{eq.4.1}
\omega_p^2 = \frac{4 \pi \alpha_{\rm em} n_{\rm e}}{m_{\rm e}} \:,
\end{equation}
where $m_{\rm e}$ and $n_{\rm e}$ are the mass and number density of the free electrons, respectively. For the present epoch halos are typically characterized by plasma densities of around $10^{-2}$ cm$^{-3}$, which means ALPs with masses such that $\frac{m_{\rm a}^2}{4}> \frac{4 \pi \alpha_{\rm em} n_{\rm e}}{m_{\rm e}}$ can undergo unsuppressed decay. This means only masses below $\lesssim 10^{-11}$ eV are excluded from decay processes by plasma effects within DM halos.

\begin{figure}[t!]
\centering
\includegraphics[width=0.65\textwidth]{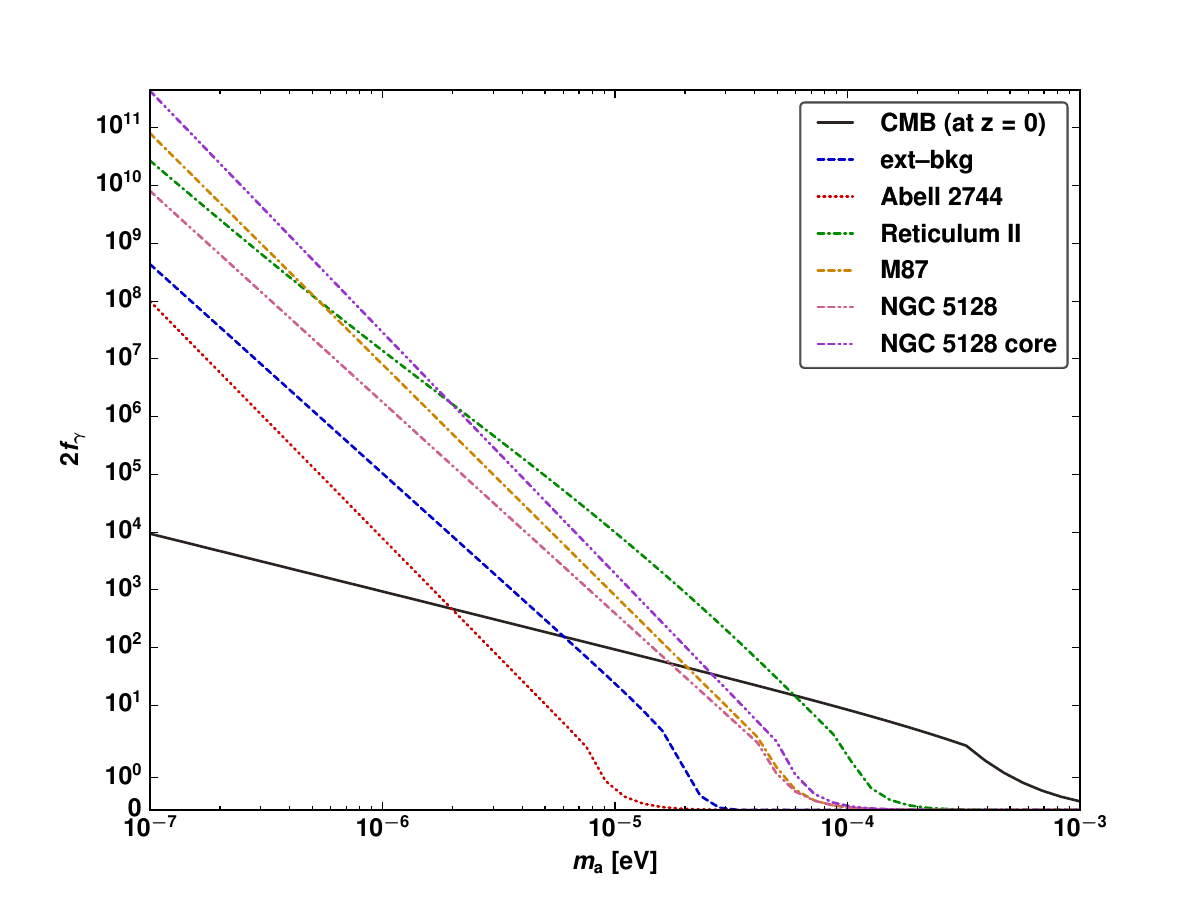}
\caption{The stimulated emission factor arising from the CMB (at $z = 0$), the extragalactic radio background, and the diffuse emission from the target environment itself.}
\label{fig.1}
\end{figure}

It remains to determine the factor $2 f _ {\gamma}$ that appears in equation~\eqref{eq.2.8}. In general, the photon occupation number $f_{\gamma}$ is a linear combination over all the sources which contribute to the photon bath with the same energy as that produced in the ALP decay. At low frequencies, we consider the photon occupation number $f_{\gamma}$ as potentially receiving contributions from the CMB, the extragalactic radio background, and the target environment diffuse emission as in reference~\cite{caputo2019detecting}
\begin{equation} \label{eq.6.1}
f_{\gamma} (m_{\rm a}) = f_{\gamma, {\rm CMB}} (m_{\rm a}) + f_{\gamma, {\rm ext\text{--}bkg}} (m_{\rm a}) + f_{\gamma, {\rm target}} (m_{\rm a}) \:,
\end{equation}
The first-order approximation to the contribution to $f_{\gamma}$ from the CMB and the extragalactic radio background can be determined from a blackbody spectrum as
\begin{equation} \label{eq.6.2}
f_{\rm \gamma, bb} (m_{\rm a})= \frac{1}{e^{\rm (E_{\gamma}/k_B T)} -1} \:,
\end{equation}
where $E_\gamma = m_{\rm a}/2$, $k_{\rm B}$ is the Boltzmann constant, and ${\rm T}$ is the blackbody temperature. The extragalactic radio background is expressed with a frequency-dependent temperature given by~\cite{fixsen2009arcade, fornengo2014isotropic}
\begin{equation} \label{eq.6.3}
{\rm T}_{\rm ext\text{--}bkg} (\nu) \simeq 1.19 \left( \frac{{\rm GHz}}{\nu} \right)^{2.62} \ {\rm K} \:,
\end{equation}
with a frequency $\nu=m_{\rm a}/4\pi$ (assuming $\hbar = 1$). While, the contribution to $f_{\gamma}$ from the galactic diffuse emission can be given by
\begin{equation} \label{eq.6.4}
f_{\gamma, {\rm gal}} (m_{\rm a})= \frac{d \rho_{\gamma}}{d E}\frac{\pi^2}{E^3_{\gamma}} \:,
\end{equation}
where $\frac{d \rho_{\gamma}}{d E}$ is the differential energy density of photons. 

In this work we consider several target halos: the galaxy cluster Abell 2744, galaxies M87 and NGC 5128, and the dwarf galaxy Reticulum II. These are all located away from the galactic plane and are observable from the SKA site. In addition, Abell 2744 and NGC 5128 host some of the southern skies' strongest diffuse radio emissions. We include M87 for comparison to \cite{caputo2019detecting} and Reticulum II as dwarf galaxies are highly DM-dominated and were explored in \cite{caputo2018looking}. For each target then, we require to know the photon density. We determine this following \cite{caputo2019detecting}
\begin{equation}
	\frac{d \rho_{\gamma}}{d E} = \frac{4\pi d_L^2 r_\gamma}{V_\gamma c} \frac{d}{d E} h\nu S(\nu)  \: ,
\end{equation} 
where $d_L$ is the luminosity distance, $S(\nu)$ is the observed flux-density spectrum, $r_\gamma$ is the radius within which most of the emission is produced, and $V_\gamma = \frac{4\pi r_\gamma^3}{3}$. This should provide an adequate estimate of the photon density within the source (note that we will confine our study of the target to within $r_\gamma$ of the halo center). For Abell 2744 we find
\begin{equation}
	S_{\rm A2744}(\nu) = 0.8 \ {\rm Jy} \ \left(\frac{\nu}{100 \ {\rm MHz}}\right)^{-1.104} \: ,
\end{equation} 
at $z \sim 0.3$, following \cite{a2744-radio} data for the full radio halo, which also extends out to $r_\gamma = 2.1$ Mpc. For M87 we use the fact that the radio luminosity between 10 MHz and 150 GHz is $9.6\times 10^{41}$ erg s$^{-1}$ emitted from within $r_\gamma = 40$ kpc and with a spectral slope of $\approx -1$~\cite{m87-radio} to determine our $f_\gamma$. 
For NGC 5218 we consider two regions, one large and extended over 1 degree (the source extension can reach up to 10 degrees but we reduce to 1 degree as this yields a physical radius of $\sim 60$ kpc, similar to M87), the other using just the core region of 6 arcminutes around the halo center. For these regions we find
\begin{align}
	S_{\rm NGC5128}(\nu) & = 7355 \ {\rm Jy} \ \left(\frac{\nu}{100 \ {\rm MHz}}\right)^{-0.655} \: , \\
	S_{\rm NGC5128,core}(\nu) & = 3365 \ {\rm Jy} \ \left(\frac{\nu}{100 \ {\rm MHz}}\right)^{-1.178} \: ,
\end{align} 
using data from \cite{ngc5128-radio} and \cite{ngc5128-core-radio1,ngc5128-core-radio2,ngc5128-core-radio3,ngc5128-core-radio4} respectively. This yields enhancement factors of the form
\begin{align}
	f_\gamma = \alpha_\gamma \left(\frac{\nu}{100 \ {\rm MHz}}\right)^{\beta_\gamma} \, \label{eq:enchament-pl}
\end{align}
with values given in table~\ref{tab:f-gamma}.
\begin{table}[ht!]
	\centering
\scalebox{0.85}{
\begin{tabular}{|p{5.166cm}|p{5.166cm}|p{5.166cm}|}
		\hline
		Target & $\alpha_\gamma$ & $\beta_\gamma$ \\
		\hline
		Abell 2744 & $8.516\times 10^3$ & $-4.104$ \\
		M87 & $8.55\times 10^6$ & $-4$ \\
		NGC 5128 & $1.78\times 10^6$ & $-3.655$ \\
		NGC 5128 core & $3.187\times 10^7$ & $-4.178$ \\
		\hline 
	\end{tabular}}
\caption{Stimulated enhancement factor parameters for various targets, see equation~(\ref{eq:enchament-pl}).}
\label{tab:f-gamma}
\end{table}

In all the targets we will include contributions from the CMB and extra-galactic radio background. In figure~\ref{fig.1}, we show the stimulated emission factor $2 f_{\gamma}$ arising from the CMB, the extragalactic radio background, and the photon field native to the various targets. 

\subsection{Magnetic conversion} \label{sec.3}
An additional process that affects ALPs is of the Primakoff type, where an ALP and a photon can be interchanged in the presence of an external magnetic field. We will estimate the contribution to radio emissions via this process by making use of the ALP-photon propagation formalism. Here we consider propagation along the $z$-axis by a beam of monochromatic photons/ALPs following~\cite{horns2012,2018axions-non-rel}
\begin{equation}
\left( -i \partial_z + \frac{1}{2 k}\mathcal{M}\right) \left( \begin{array}{c} A_{x} \\ A_{y} \\ a \end{array} \right) = 0 \: , \label{eq:lin}
\end{equation}
where $k^2 = E^2 - m_a^2$, $A_{x}$ and $A_y$ are the polarisation states, and $a$ is the ALP field.
Here $E$ and $\omega$ are interchanged under the assumption of natural units. The matrix $\mathcal{M}$ is now of the form
\begin{equation}
	\mathcal{M} = \left( \begin{array}{ccc} -\omega_{pl}^2 & 0 & g_{a\gamma} B_x m_a  \\
		0 & -\omega_{pl}^2 & g_{a\gamma} B_y m_a \\
		g_{a\gamma} B_x m_a & g_{a\gamma} B_y m_a & -m_a^2\\ \end{array} \right) \: ,
	\label{eq:m2}
\end{equation}
where $\omega_{\rm pl}^2 = 4 \pi \alpha \frac{n_e}{m_e}$ is the plasma frequency for number density $n_e$, $g_{a\gamma}$ is the ALP-photon coupling, $E$ is the particle energy, $m_a$ is the ALP mass, and $B_i$ is the component of the magnetic field in direction $i$.

In this formulation we have neglected the Faraday rotation terms. This is justified purely in terms of simplicity as it greatly simplifies the diagonalization of equation~\eqref{eq:m2}. Additionally, we will be considering multiple magnetic domains with random field orientations which ensures even mixing between ALP and both polarization states (this should mitigate the omission of the Faraday terms).

In order to calculate the effect of ALP-photon mixing with multiple magnetic domains, we sample domain lengths from a distribution given by the magnetic field power spectrum and assign a random magnetic field orientation within each domain. We then solve equation~(\ref{eq:lin}) in each domain to obtain the fraction of ALPs converted into photons. We compute this for a variety of paths through the target environment, each defined by an impact parameter. Each has $p_{a\to\gamma}$ averaged over 10000 realizations of the random magnetic field configuration. This becomes a conversion rate, $\Gamma_{\rm conv}$, when dividing by the average time required to follow the path considered (this being $\frac{l}{\sigma_v}$ where $l$ is the path length). 

We will use M31 as a test case, as there is data on the gas and magnetic field distributions available~\cite{ruiz-granados2010}. This yields
\begin{align}
	B_{\rm M31} & = \frac{984 \ {\rm kpc}}{200 \ {\rm kpc} + r} \ \mu {\rm G} \: , \\
	n_{\rm M31} & = 0.06 \exp{\left(-\frac{r}{5 \ {\rm kpc}}\right)} \ {\rm cm}^{-3} \: , \\
	\rho_{\rm M31} & = \frac{1.09\times 10^7}{\left(\frac{r}{63 \ {\rm kpc}}\right)\left(1+ \frac{r}{63 \ {\rm kpc}}\right)^2} \ {\rm M}_\odot \ {\rm kpc}^{-3} \: , 
\end{align}
where $\rho_{\rm M31}$ is drawn from~\cite{tamm2012stellar}. We will assume the magnetic field is turbulent with a coherence length of $0.1$ kpc and a power-spectral index of $5/3$.  We will also use Reticulum II, whose data in this regard are more uncertain~\cite{regis2017dark}
\begin{align}
	B_{\rm Ret. II} & = 1 \ \mu {\rm G} \ \exp{\left(-\frac{r}{35 \ {\rm pc}}\right)} \: , \\
	n_{\rm Ret. II} & = 1\times 10^{-6} \exp{\left(-\frac{r}{35 \ {\rm pc}}\right)} \ {\rm cm}^{-3} \:. 
\end{align}

\section{Surface brightness of ALP decays/conversions} \label{sec.4}
Given a DM halo constructed of ALP condensates, we can estimate the surface brightness from decays at impact radius $R$ to be
\begin{equation} \label{eq:sb-alp}
	I_{\rm dec} (R) = \frac{{\rm \Gamma}_{\rm eff}}{4 \pi} \frac{\sigma_v}{c} \int_{-\infty}^{\infty} d l \,  \rho_{\rm a}(r) \: ,
\end{equation}
where $l$ is the coordinate along the line of sight, $r = \sqrt{l^2 + R^2}$ is the spherical radius, $R$ is the distance to the halo center, and $\rho_{\rm a}$ is the ALP density. Thus, the signal would appear as a narrow spectral line, broadened by the ALPs velocity dispersion $\sigma_v$, centered on a frequency corresponding to $m_{\rm a}/2$. For magnetic conversion we have 
\begin{equation}
	I_{\rm conv} = \frac{\Gamma_{\rm conv}}{4\pi} \frac{\sigma_v}{c} \int_{-\infty}^{\infty} d l \, \rho_{\rm a}(r) \: ,
\end{equation} 
where the signal will be in the form of a narrow line centered at frequency $\nu = \frac{m_a}{4\pi}$, broadened by $\sigma_v$. We will consider the time for ALPs to move a distance $l$ as $\frac{l}{\sigma_v}$.

Thus, in order to determine this value we need the DM density profile for a halo of interest. As noted previously we will be considering the galaxy cluster Abell 2744, galaxies M87 and NGC 5128, as well as the dwarf galaxy Reticulum II. We will be making use of NFW~\cite{nfw1996}, cgNFW~\cite{m87-virial}, and Einasto~\cite{einasto1968} profiles
\begin{align}
	\rho_{{\rm NFW}}(r) & = \frac{\rho_s}{\frac{r}{r_{\rm s}} \left(1 + \frac{r}{r_{\rm s}}\right)^2} \: , \\
	\rho_{\rm cgNFW}(r) & = \rho_{\rm s} \left(\frac{r+r_{\rm c}}{r_{\rm s}}\right)^{-\gamma} \left(1 + \frac{r}{r_s}\right)^{\gamma-3} \: , \\
	\rho_{{\rm e}}(r) & =\rho_{\rm s} \exp\left[-\frac{2}{\alpha} \left(\left[\frac{r}{r_{\rm s}}\right]^{\alpha} - 1\right)\right] \: ,
\end{align}
where $\rho_{\rm s}$ is the characteristic density, $r_{\rm s}$ the characteristic radius, $r_{\rm c}$ the core radius, $\gamma$ the slope index, and $\alpha$ the einasto shape parameter.
The parameters used are listed in table~\ref{tab:halos}.
\begin{table}[ht!]
	\centering
	\begin{tabular}{|l|l|l|l|l|l|}
		\hline
		Halo & Profile & $\rho_{\rm s}$ (M$_\odot$ kpc$^{-3}$) & $r_s$ (kpc) & $\sigma_{\rm v}$ (km s$^{-1}$) & $r_{\rm c}$ (kpc) \\
		\hline
		Abell 2744 & NFW & $1.54 \times 10^5$~\cite{a2744-virial} & 1380~\cite{a2744-virial} & $10^3$ & - \\
		M87 & cgNFW ($\gamma=2.54$) & $3.98 \times 10^5$~\cite{m87-virial} & $273$~\cite{m87-virial} & 383~\cite{m87-sigmav} & $63$~\cite{m87-virial}  \\
		NGC 5128 & NFW & $5.76\times 10^6$ & $27.9$ & 130~\cite{ngc5128-sigmav} & - \\
		Reticulum II & Einasto ($\alpha=0.4$) & $7\times 10^7$~\cite{regis2017dark} & $0.2$~\cite{regis2017dark} & $4$~\cite{koposov2015,bechtol2015} & - \\
		\hline
	\end{tabular}
	\caption{The DM density profiles in use for the target halos. Note that for Abell 2744 we assume a velocity dispersion similar to the Coma cluster~\cite{lokas2003dark}. For NGC 5128 we infer the properties from those given in \cite{ngc5128-virial} with the concentration relation from \cite{prada2014}.}
\label{tab:halos}
\end{table}



\section{MeerKAT and the SKA} \label{sec.5}
In recent years, there is increasing interest in using radio telescopes to search for a radio signal produced by cold dark matter. As when completed, the SKA will be the largest radio telescope built on earth, so it is considered to be the most sensitive radio telescope ever~\cite{colafrancesco2015probing}. This makes it the radio astronomy frontier in the hunt for any indirect signature of dark matter. The MeerKAT radio telescope is a precursor to the full SKA system and it will be integrated into the first phase of the full SKA~\cite{booth2012overview, pourtsidou2016prospects}. A simple estimate of sensitivity of radio telescopes can be determined via the minimum detectable temperature~\cite{radioAstronFund}
\begin{equation}
	T_{{\rm min}} = \frac{T_{{\rm sys}}}{\sqrt{\Delta B t_{{\rm obs}} N }} \: ,
\end{equation}
where $T_{{\rm sys}}$ is the total noise from both the background and antenna, $\Delta B$ is the bandwidth, $t_{{\rm obs}}$ is the observing time, and $N$ is the number of antennae being employed. By using the relationship between incoming flux $S$ and antenna temperature $T$:
\begin{equation}
	T = \frac{S A_{{\rm eff}}}{2} \: ,
\end{equation} 
where $A_{{\rm eff}}$ is the effective collecting area for the telescope. Combining these expressions allows us to find the minimum detectable flux
\begin{equation}
	S_{{\rm min}} = 2\frac{T_{{\rm sys}}}{A_{{\rm eff}}\sqrt{\Delta B t_{{\rm obs}} N }} \: . \label{eq:sens}
\end{equation}

In previous works this has been used with SKA baseline design parameters\footnote{\url{https://www.skatelescope.org/wp-content/uploads/2013/08/ska-tel-sko-dd-001-1_baselinedesign1.pdf}} combined with a point-source assumption to provide a sensitivity figure. To account for frequency-dependent degradation effects we will make use of the results from tables 6 and 7 of \cite{braun2019anticipated}. In addition, we will make use of surface brightness calculations to avoid point-source assumptions. Hence, we will be making use of surface brightness sensitivities, in units of Jy/arcminute$^2$. These are derived from the Jy/beam figures quoted in \cite{braun2019anticipated} via the beam size and parameters of equation~\eqref{eq:sens}. This includes line emission sensitivity estimates for a width $\frac{\Delta \nu}{\nu} = 10^{-4}$, which we will rescale to the line width of ALP decay in a given halo via $\frac{\Delta \nu}{\nu} = \frac{\sigma_v}{c}$, where $\sigma_v$ is the velocity dispersion of the target DM halo. In the case of MeerKAT we employ SARAO's publicly available tools~\footnote{\url{https://skaafrica.atlassian.net/wiki/spaces/ESDKB/pages/41091107/Sensitivity+calculators}}.

A further point we will consider is the use of a visibility taper. This down-weights the contributions of the long-baselines in an interferometer. The effect is to reduce sensitivity to small-scale emissions~\cite{radioAstronFund}. We will consider a tapering scale of 15 arcseconds, following the demonstration of the effectiveness for WIMP searches in~\cite{regis2017dark} in nearby dwarf galaxies.
In our analysis of visibility tapers, we determine the MeerKAT sensitivities via the Stimela~\cite{makhathini2018advanced} software package. This allows us to find the rms image noise after simulating MeerKAT observations with the NVSS source catalog as a sky model~\cite{condon1998nrao}. To do this, Stimela makes use of CASA~\cite{mcmullin2007casa} for observation simulation, RFIMasker\footnote{\url{https://github.com/bennahugo/RFIMasker}} to mask out frequency channels with static interference, Meqtrees~\cite{noordam2010meqtrees} for calibration and visibility simulation, and WSClean~\cite{offringa2014wsclean} for imaging (at robust weighting $1$). All our sensitivities apply to the MeerKAT L-band (890 MHz to 1.65 GHz) and assumed a channel width of 1 MHz for the simulation of continuum observations. The ratio of sensitivity and beam-size changes for MeerKAT continuum simulations are then assumed to apply line emissions as well as the SKA.

\section{Results} \label{sec.6} 

For the case of magnetic conversion we note that, in dwarf galaxies like Reticulum II we find that, for $m_a = 10^{-7}$ eV and $g_{a\gamma} \sim 10^{-10}$ GeV$^{-1}$, we have that $f_{a\to\gamma} \lesssim 10^{-35}$ (across the whole SKA frequency range) which yields a conversion rate of around $10^{-51}$ s$^{-1}$. This is much slower than even the perturbative decay time for this ALP mass. In the case of a galaxy like M31, these rates are even slower. Thus, we conclude that there is no significant contribution to radio emissions from magnetic conversion. The surprising weakness of this process is down to the similarity of the ALP mass and energy when considered as part of a CDM halo. This means that $\frac{m_a^2}{k}$ is large and the conversion process, from ALP to photon, is thereby suppressed. We note, however, that more powerful radio emissions are obtainable in other environments, such as near neutron stars~\cite{2018axions-non-rel,witte2021}.

From figure~\ref{fig.1}, it is very clear that the decay rate of ALPs with masses $m_{\rm a} \gtrsim 10^{-4} \ {\rm eV}$ does not receive a remarkable enhancement. Therefore, in this case, it is not expected that the stimulated decay has an essential role, and we cannot count on the spontaneous decay of ALPs to produce a significant observational signal because of the low rate of this process. Consequently, this can be considered as an upper limit for the ALP mass that can be accountable for any significant radio signature produced from the stimulated decay of ALPs. In addition, the typical plasma density in galactic halos (in the present epoch), where dark matter is expected to be highly concentrated, can prevent the stimulated decay of ALPs with masses $m_{\rm a} \lesssim 10^{-11} \ {\rm eV}$. This puts a lower limit on the ALP mass at which the stimulated decay of ALPs is allowed by the plasma effects. Hence, with the current-epoch plasma density, the stimulated decay of ALP with the $10^{-11} \ \text{--} \ 10^{-4} \ {\rm eV}$ mass range seems to be effective and allowed by the plasma effects within DM halos.

Within this range, the search for ALPs in the $10^{-7} \ \text{--} \ 10^{-4} \ {\rm eV}$ mass range seems to be the most exciting scenario due to the sensitivity limitations of the current and next-generation radio telescopes. In this case, we would expect the stimulated decay of ALPs with this mass range to produce low-frequency radio emission at frequencies from $60 \ {\rm MHz}$ to $12 \ {\rm GHz}$. Considering that the SKA and MeerKAT telescopes will operate over frequencies range from $50 \ {\rm MHz}$ to $20 \ {\rm GHz}$~\cite{cembranos2020ska} and 890 MHz to 1.65 GHz, this implies that they would be able to detect photons produced from the decay of ALPs with mass ranges $4.96 \times 10^{-7} \ \text{--} \ 1.04 \times 10^{-4} \ {\rm eV}$ and $4.80 \times 10^{-6} \ \text{--} \ 1.36 \times 10^{-5} \ {\rm eV}$ respectively. This agrees with the suggestion in~\cite{caputo2019detecting} to search for a radio signal based of the stimulated decay of axion dark matter in the $10^{-7} \ \text{--} \ 10^{-3} \ {\rm eV}$ mass range in the near-future radio observations by SKA or using forthcoming axion search experiments, such as ALPS-II~\cite{bahre2013any} and IAXO~\cite{armengaud2019physics}.

\begin{figure}[t!]
\centering
\includegraphics[width=0.495\textwidth]{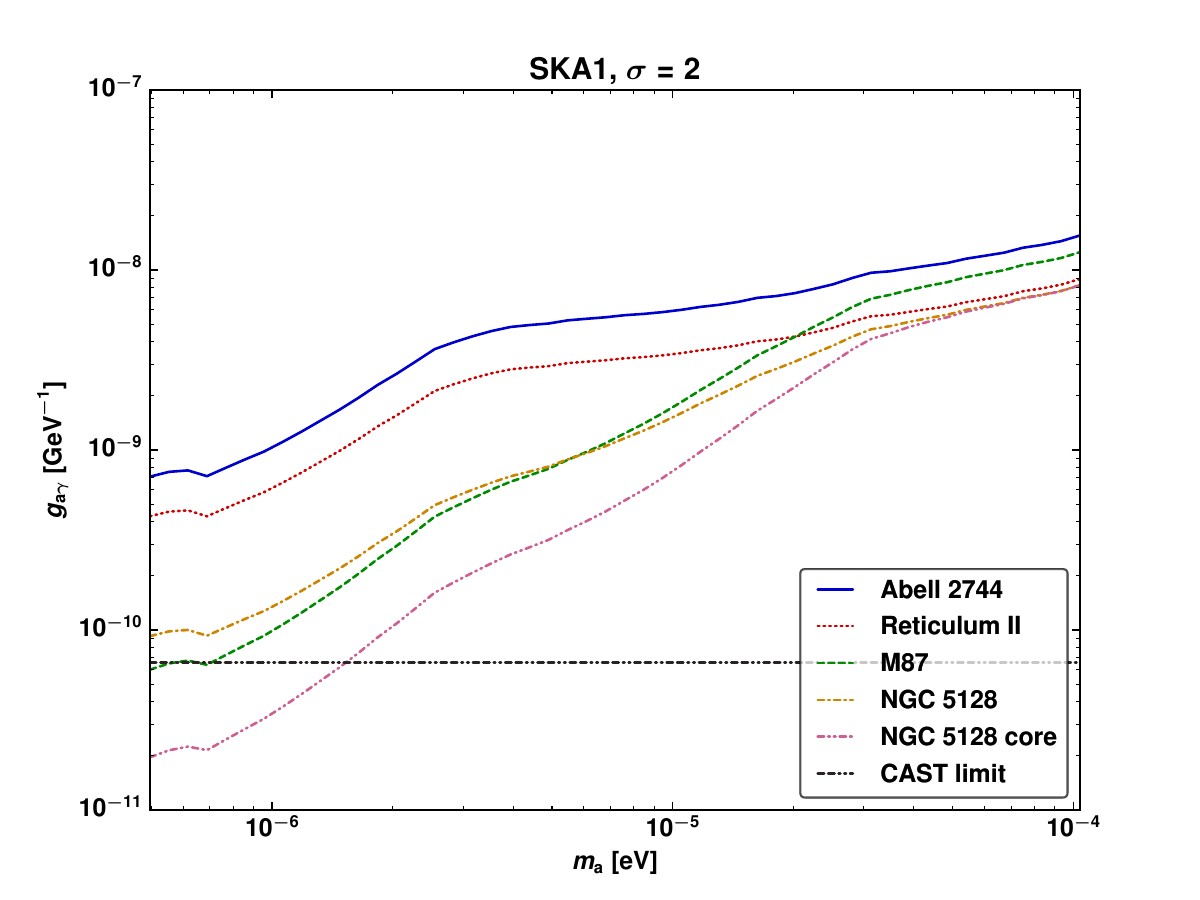}
\includegraphics[width=0.495\textwidth]{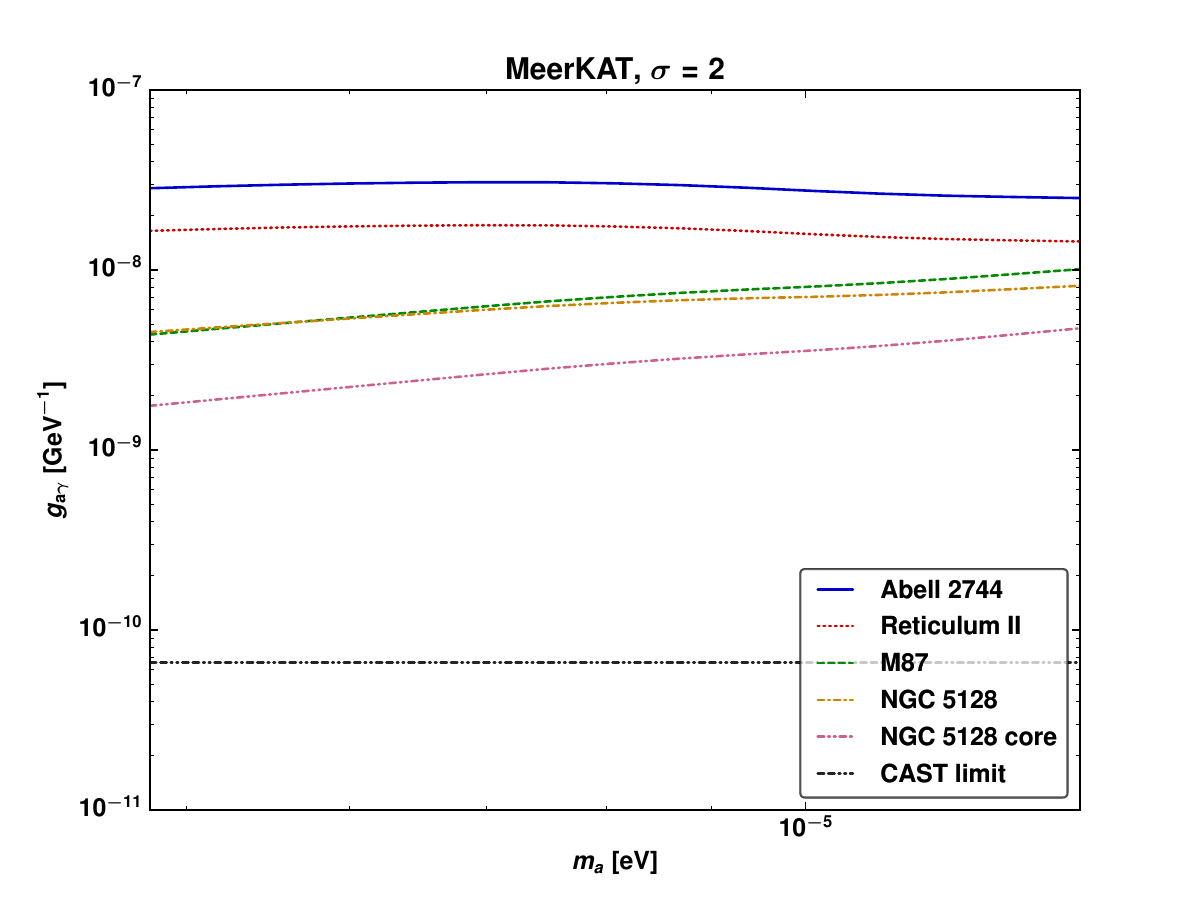}
\caption{The potential non-observation constraints with SKA1 and MeerKAT telescopes for 2$\sigma$ confidence intervals using halo surface brightness from ALP decay. Sensitivities are untapered and taken directly from sources cited in the text.}
\label{fig.3}
\end{figure}

In figure~\ref{fig.3}, we illustrate the potential non-observation constraints with SKA (phase 1) and MeerKAT telescopes, for 2$\sigma$ confidence intervals, using line sensitivities within 50 hours of observing time. Note that, the use of the 2$\sigma$ confidence interval allows us to estimate the exclusion limits in the event of non-observation. In table~\ref{tab.3}, we summarize the lower limits of the ALP-photon coupling $g_{a\gamma}$ that can be detectable by the SKA and MeerKAT radio telescopes with the 2$\sigma$ confidence intervals for the most promising case of the NGC 5218 core region. These values are in the range of $1.96 \times 10^{-11} \ \text{--} \ 8.23 \times 10^{-9} \ {\rm GeV}^{-1}$ for ALP mass in the range of $4.96 \times 10^{-7}\ \text{--} \ 1.04 \times 10^{-4} \ {\rm eV}$ for the SKA telescope. While in the case of the MeerKAT telescope, the limits are in the range of $1.76 \times 10^{-9} \ \text{--} \ 4.75 \times 10^{-9} \ {\rm GeV}^{-1}$ for ALP mass in the range of $4.80 \times 10^{-6}\ \text{--} \ 1.36 \times 10^{-5} \ {\rm eV}$.

For SKA, these potential limits are generally weaker (for $m_a > 4\times 10^{-6}$ eV) than existing CAST results, i.e. $g_{a\gamma} \lesssim 6.6 \times 10^{-11} \ {\rm GeV}^{-1}$~\cite{anastassopoulos2017new}. Additionally, they are also weaker than projections for the ALPS-II experiment, i.e. $g_{a\gamma} \lesssim {\rm few} \times 10^{-11} \ {\rm GeV}^{-1}$~\cite{bahre2013any}, and the IAXO experiment, i.e. $g_{a\gamma} \lesssim {\rm few} \times 10^{-12} \ {\rm GeV}^{-1}$~\cite{armengaud2019physics}. The only point of competition with these experiments comes in at the low end of the mass range.
MeerKAT, however, can only probe a region in the ALP mass-coupling parameter space that has already been ruled out by the CAST experiment. In comparison to the related works~\cite{caputo2019detecting, caputo2018looking}, which found SKA (phase 1) limits of $g_{\rm \gamma a} \simeq 8 \times 10^{-11} \ \text{--} \ 3 \times 10^{-10} \ {\rm GeV}^{-1}$, we see our NGC 5128 limits slightly improve on this at low frequencies but depend far more strongly on ALP mass. However, if we compare our dwarf galaxy limits we find our results are an order of magnitude weaker (and still include stronger frequency dependence). This suggests that sensitivity estimates in~\cite{caputo2018looking,caputo2019detecting} were over-optimistic and neglected important frequency-dependent effects included in~\cite{braun2019anticipated}. We note that NGC 5128 core produces the most promising results due to the strong galactic background photon field, as this provides the dominant contribution to the Bose enhancement.

Our results indicate that the untapered sensitivity of the SKA and the MeerKAT telescopes has a limited potential for probing the ALP coupling-mass parameter space, excepting at low ALP masses $\lesssim 4 \times 10^{-6}$ where the SKA can produce some competitive results. To improve these limits, we consider the effects of a visibility taper on the telescope sensitivities. This is motivated by the fact that emissions from ALPs in a dark matter halo will be diffuse. Thus, a visibility taper, that reduces the contributions from long array baselines (small scales) will effectively tune the telescope to focus on larger scale emissions.

\begin{figure}[t!]
\centering
\includegraphics[width=0.495\textwidth]{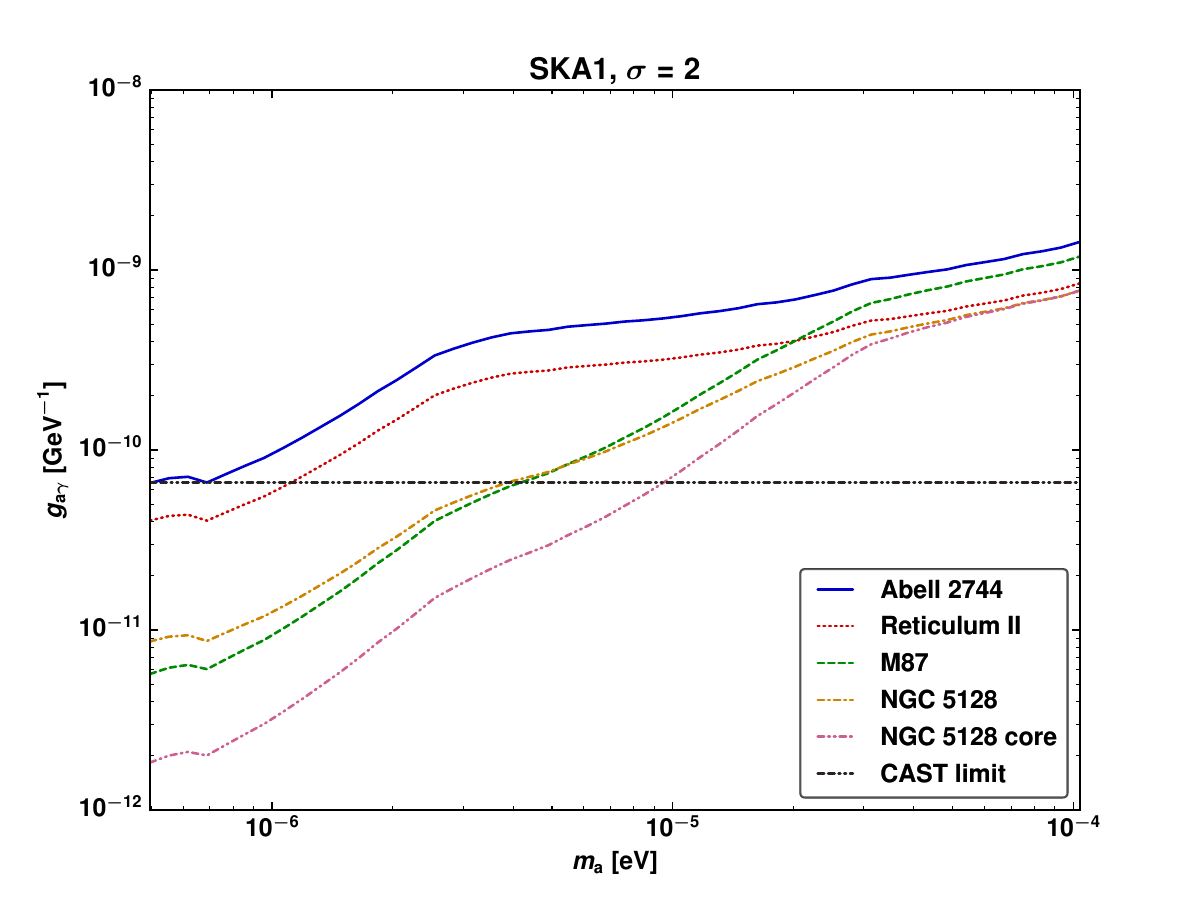}
\includegraphics[width=0.495\textwidth]{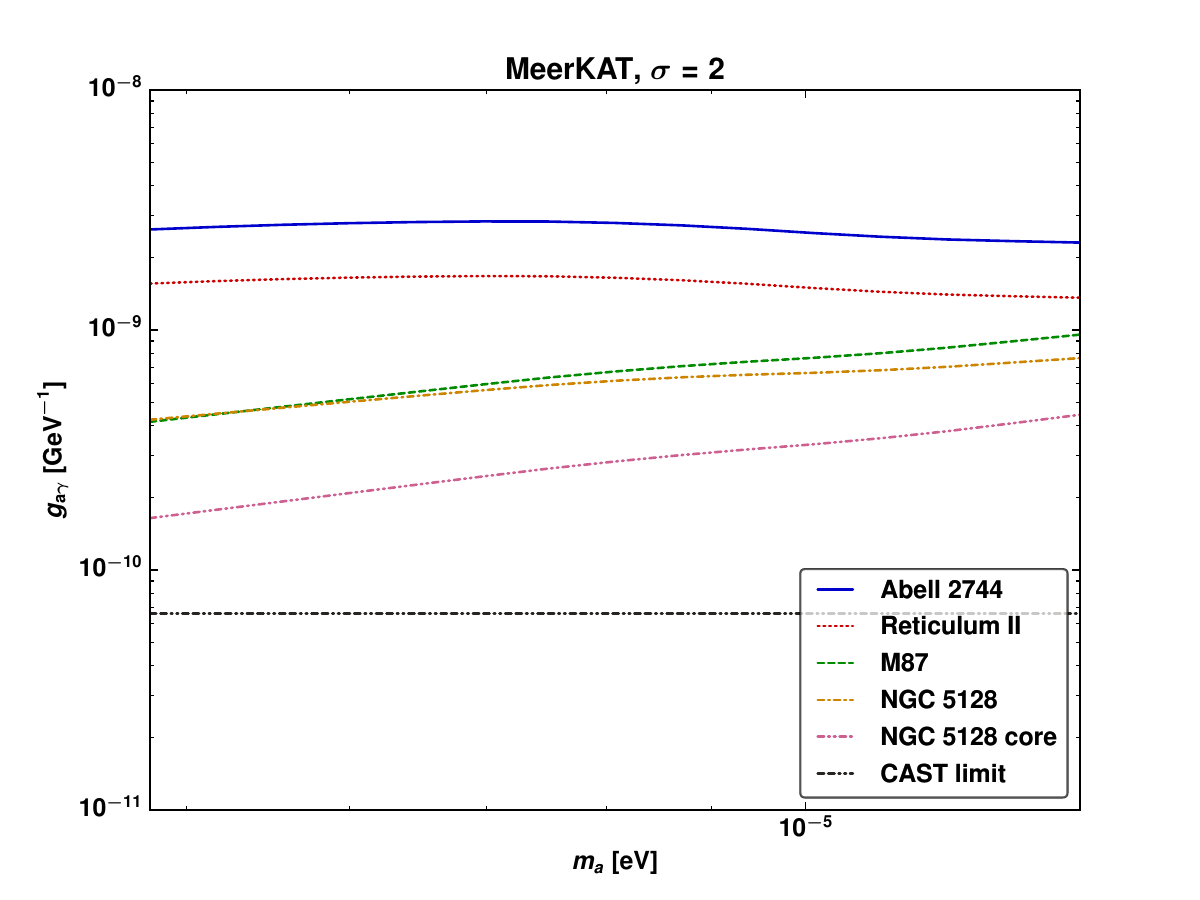}
\caption{The potential non-observation constraints with SKA1 and MeerKAT telescopes for 2$\sigma$ confidence intervals using halo surface brightness from ALP decay. In this plot, a 15 arcsecond taper on the visibilities is employed (see text). The CAST limit is taken from~\cite{anastassopoulos2017new}.}
\label{fig.4}
\end{figure}

In figure~\ref{fig.4}, we display the effect on SKA and MeerKAT sensitivity for 2$\sigma$ confidence intervals using a 15 arcsecond taper on visibilities. The taper provides an improvement on the limits by around an order of magnitude across the studied mass ranges compared to the limits we obtained without applying the tapering effects. The results for the 15 arcsecond taper in the most promising case of the NGC 5128 core region are summarized in table~\ref{tab.4}. Now the MeerKAT telescope is expected to reach limits on the ALP-photon coupling of $1.65 \times 10^{-10} \ \text{--} \ 4.44 \times 10^{-10} \ {\rm GeV}^{-1}$ for ALP mass in the range of $4.80 \times 10^{-6} \ \text{--} \ 1.36 \times 10^{-5} \ {\rm eV}$. These values are competitive with the projected limits in~\cite{caputo2019detecting} and the sensitivity reach of the CAST experiment. On the other hand, the SKA telescope is able to cover a sensitivity region of the ALP parametric space in the range of $1.83 \times 10^{-12} \ \text{--} \ 7.69 \times 10^{-10} \ {\rm GeV}^{-1}$ for ALP mass in the range of $4.96 \times 10^{-7} \ \text{--} \ 1.04 \times 10^{-4} \ {\rm eV}$. The projected limits in the SKA case now exceed the suggested limits in~\cite{caputo2019detecting} and the sensitivity reach of the CAST experiment by more than one order of magnitude and overlap with the potential sensitivity region for the IAXO experiment with ALP masses $m_{\rm a} \lesssim 2 \times 10^{-6} \ {\rm eV}$. Thus, by tailoring radio observations to the observation of diffuse emissions we can provide radio probes complementary with experiments like ALPS-II and IAXO. Without these measures the accurate SKA and MeerKAT sensitivities show limited potential to the constrain these important regions of the ALP parameter space.
\begin{table}[ht!]
\centering
\scalebox{0.85}{
\begin{tabular}{|p{3.1cm}||p{3.1cm}|p{3.1cm}||p{3.1cm}|p{3.1cm}|}
 \hline
\multirow{2}{*}{Confidence interval} & \multicolumn{2}{|c||}{SKA1} & \multicolumn{2}{|c|}{MeerKAT}\\
\cline{2-5}
&$m_{\rm a} \ [{\rm eV}]$& $g_{a\gamma} \ [{\rm GeV}]^{-1}$&$m_{\rm a} \ [{\rm eV}]$&$g_{a\gamma} \ [{\rm GeV}]^{-1}$ \\
 \hline \hline
\multirow{2}{*}{$2\sigma$} & $1.04 \times 10^{-4}$&$8.23 \times 10^{-9}$&$1.36 \times 10^{-5}$&$4.75 \times 10^{-9}$\\ 
\cline{2-5}
 &$4.96 \times 10^{-7}$&$1.96 \times 10^{-11}$&$4.80 \times 10^{-6}$&$1.76 \times 10^{-9}$\\ 
 \hline 
\end{tabular}}
\caption{Upper limits of the ALP to photon coupling $g_{a\gamma}$ that can be probed by SKA and MeerKAT telescopes with the $2 \sigma$ confidence intervals for different ALP mass categories and using untapered sensitivities. These limits obtained from the most promising case of the NGC 5128 core.}
\label{tab.3} 
\end{table}

\begin{table}[ht!]
\centering
\scalebox{0.85}{
\begin{tabular}{|p{3.1cm}||p{3.1cm}|p{3.1cm}||p{3.1cm}|p{3.1cm}|}
 \hline
\multirow{2}{*}{Confidence interval} & \multicolumn{2}{|c||}{SKA1} & \multicolumn{2}{|c|}{MeerKAT}\\
\cline{2-5}
&$m_{\rm a} \ [{\rm eV}]$& $g_{a\gamma} \ [{\rm GeV}]^{-1}$&$m_{\rm a} \ [{\rm eV}]$&$g_{a\gamma} \ [{\rm GeV}]^{-1}$ \\
 \hline \hline
\multirow{2}{*}{$2\sigma$} & $1.04 \times 10^{-4}$&$7.69 \times 10^{-10}$&$1.36 \times 10^{-5}$&$4.44 \times 10^{-10}$\\ 
\cline{2-5}
 &$4.96 \times 10^{-7}$&$1.83 \times 10^{-12}$&$4.80 \times 10^{-6}$&$1.65 \times 10^{-10}$\\
 \hline 
\end{tabular}}
\caption{Upper limits of the ALP to photon coupling $g_{a\gamma}$ that can be probed by SKA1 and MeerKAT telescopes with the $2 \sigma$ confidence intervals for different ALP mass categories and using a 15 arcsecond taper on the visibilities. These limits obtained from the most promising case of the NGC 5128 core.}
\label{tab.4} 
\end{table}

\section{Conclusion} \label{sec.7}

Axion-like particles are a very well-motivated candidate to account for the cold dark matter content in the universe. In this work, we studied the possibility of detecting an observable signature produced due to the decay of CDM ALPs into photons at radio frequencies. For ALPs with masses and coupling with photons allowed by astrophysical and laboratory constraints, the ALPs are very stable on the cosmological scale, and their spontaneous decay is unlikely to be responsible for producing any detectable radio signal. Since ALPs are identical bosons, they may form a Bose-Einstein condensate with very high occupation numbers. The presence of an ambient radiation field leads to a stimulated enhancement of the decay rate. Depending on the astrophysical environment and the ALP mass, the stimulated decay of ALPs BEC in an expanding universe and under the plasma effects can be counted for producing observable radio signals enhanced by several orders of magnitude.

The stimulated decay of ALPs arising from the presence of the ambient background of photons from the CMB, the extragalactic radio background, and the diffuse emission from target environment itself can result in a large enhancement of related emissions. However, the decay rate of ALPs with masses $\sim 10^{-4}$ eV does not receive a remarkable enhancement from these sources. This puts an upper limit for the ALP mass that can be subject to stimulated decay. In addition, the plasma effects within DM halos and and the current sensitivity of the radio telescopes set a lower bound on the ALP mass $\sim 10^{-7}$ eV that can be accountable for any significant radio signature. Previous work, such as~\cite{caputo2018looking,caputo2019detecting} had found potential non-observation limits with SKA (phase 1) that were relatively independent of ALP mass (in the $10^{-7}$ to $10^{-4}$ eV mass range). By using more realistic sensitivity estimates, corrected for frequency-dependent degradation, from~\cite{braun2019anticipated}, we were able to show that the limits would actually have a power-law dependence on ALP mass with index $~1$. In addition, our dwarf galaxy limits were at least an order of magnitude weaker than those reported in~\cite{caputo2018looking}. It is notable that our best limits were found in the core region of the NGC 5128 galaxy, due to the strong photon field present within radio galaxies. This enhancement due to the ambient photons outstripped even the normal advantages, of density and nearness, ascribed to dwarf galaxies. Thus, we showed that SKA could complement ALPS-II, via radio galaxy observations, but not IAXO. MeerKAT, however, could only probe regions of the parameter space already excluded by CAST. It is notable that our comparison to telescope sensitivities is likely also more accurate as we used a surface brightness approach (this being directly comparable to sensitivities quoted in the conventional Jy/beam units) rather then integrated fluxes that rely on point source assumptions.

We also considered the scenario where the observations were tailored to hunting diffuse emissions via the use of a 15 arcsecond visibility taper. This enhanced the sensitivity by an order of magnitude. With ALPs in the mass range of $4.96 \times 10^{-7} \ \text{--} \ 1.04 \times 10^{-4} \ {\rm eV}$ being limited such that $1.83 \times 10^{-12} \ \text{--} \ 7.69 \times 10^{-10} \ {\rm GeV}^{-1}$ for the SKA. This would allow SKA to compete with the IAXO experiment using 50 hours of observation time on the core of the NGC 5128 galaxy.

Finally, we determined that magnetic conversion of ALPs to photons, in galactic halos, is at least 5 orders of magnitude slower than even spontaneous decay. This confirms suggestions from previous works~\cite{caputo2019detecting} that made use of more approximate formulations of the conversion process. 

\acknowledgments

This work is based on the research supported by the South African Research Chairs Initiative of the Department of Science and Technology and National Research Foundation of South Africa (Grant No 77948). A. Ayad acknowledges support from the Department of Science and Innovation/National Research Foundation (DSI/NRF) Square Kilometre Array (SKA) post-graduate bursary initiative under the same Grant. G. Beck acknowledges support from a National Research Foundation of South Africa Thuthuka grant no. 117969. The authors would like also to offer special thanks to Prof. S. Colafrancesco, who, although no longer with us, continues to inspire by his example and dedication to the students he served over the course of his career.

\bibliographystyle{JHEP}
\bibliography{references}

\end{document}